\documentclass[10pt]{iopart}

\usepackage{iopams}

\expandafter\let\csname equation*\endcsname\relax
\expandafter\let\csname endequation*\endcsname\relax

\usepackage{mathptmx}
\usepackage{amsmath}
\usepackage{graphicx}

\graphicspath{{../common-fig/}{fig/}}

\renewcommand\fbox

\newcommand{\tramp}{0.75}
\newcommand{\tlate}{2.00}
\newcommand{\tflat}{9.50}
\newcommand{\ipramp}{0.8}
\newcommand{\iplate}{2.0}
\newcommand{\ipflat}{8.7}
\newcommand{\vscramp}{13.2}
\newcommand{\vsclate}{3.2}
\newcommand{\vscflat}{30.3}
\newcommand{\vesramp}{52.5}
\newcommand{\veslate}{48.0}
\newcommand{\vesflat}{6.1}
\newcommand{\vspramp}{10.2}
\newcommand{\vsplate}{7.9}
\newcommand{\vspflat}{0.9}
\newcommand{\rmseezpzramp}{1.31}
\newcommand{\rmseeupzramp}{1.18}
\newcommand{\rmseezpzlate}{0.57}
\newcommand{\rmseezpulate}{0.56}
\newcommand{\rmseeupzlate}{0.47}
\newcommand{\rmseeupulate}{0.41}
\newcommand{\rmseezpzflat}{0.21}
\newcommand{\rmseezpuflat}{0.22}
\newcommand{\rmseeupzflat}{0.27}
\newcommand{\rmseeupuflat}{0.15}
\newcommand{\maaeezpzramp}{3.63}

\newcommand{\maaeezpulate}{1.11}

\newcommand{\maaeezpuflat}{0.76}

\newcommand{\maaelatemin}{0.27}
\newcommand{\maaelatemax}{2.62}
\newcommand{\rmsefullmax}{1.96}
\newcommand{\maaefullmax}{5.68}

\begin{document}

\title[GSA for the SPARC PRD]{Gauss' Separation Algorithm for the Magnetic Field Decomposition of the SPARC PRD simulation}
\author{Gregorio L. Trevisan$^1$, Ryan M. Sweeney$^{1,2}$, Robert S. Granetz$^1$}
\address{$^1$ MIT Plasma Science and Fusion Center, Cambridge MA, USA}
\address{$^2$ Commonwealth Fusion Systems, Cambridge MA, USA}
\ead{gtrevisan@psfc.mit.edu}

\date{\today}

\begin{abstract}
A magnetic separation algorithm originally introduced by Gauss has been revisited in recent years with application to magnetically-confined fusion experiments.
The main result offered by Gauss' Separation Algorithm~(GSA) is a magnetic field computation that enables the decomposition of the contributions due to sources internal and external to a bounded volume through the generalized Virtual-Casing Principle.
The present work applies GSA to simulations of the SPARC Primary Reference Discharge and demonstrates its separation capability on all phases of the shot, including the ramp-up during which significant currents on the passive structures often complicate the reconstruction of plasma equilibria.
Such early results suggest a promising new way to study ramp-ups, disruptions, and vertical displacement events without any specific modeling assumption.
Envisaged applications include reading magnetic data from experimental diagnostics and producing a new set of signals without external contributions to be used by downstream equilibrium codes for improved accuracy or speed.
\end{abstract}


\vspace{2em}
\noindent{\it Original submission\/}: September 2023

\noindent{\it Final revision\/}: May 2024

\noindent{\it arXiv submission\/}: July 2025

\maketitle

\ioptwocol

\section{Introduction}

The so-called Gauss' Separation Algorithm~(GSA), first introduced by Gauss in 1839 while tackling terrestrial magnetism~\cite{gauss1877, glassmeier2014} and later exploited by many branches of research including planetary sciences, has been revisited in recent years and applied to plasma physics and tokamak experiments.
Most notably, Sweeney and Strait tackled the problem using a simplified cylindrical geometry to arrive at semi-analytical expressions for the components of the magnetic field which leveraged Bessel functions, studied rotating tearing modes on DIII-D~\cite{sweeney2019}, and later applied the decomposition to electromagnetic torque measurements during locked modes~\cite{strait2021}.
Additionally, Plessers and Artola extended the concept to a toroidal geometry and leveraged the non-linear reduced magnetohydrodynamic~(MHD) code JOREK/STARWALL to apply the magnetic separation to simulated ITER vertical displacement events~\cite{plessers2020}.
In this work we introduce a novel and general code, named \texttt{GaSATO}, to numerically perform the GSA separation on the simulations of the SPARC Primary Reference Discharge~(PRD)~\cite{creely2020}.

Given that GSA as a magnetic separation algorithm relies on the Virtual-Casing Principle~(VCP), significant overlap exists with research involving the VCP and its applications in tokamak physics.
The VCP was originally introduced by Shafranov and Zakharov, who remarked the relationship between the normal components of the external (``vacuum'') and internal (``plasma'') magnetic field on a superconducting surface separating the two regions, and offered a simplified approach for the computation of the former~\cite{shafranov1972}.
While most often the separating surface was chosen to be a flux surface, further work included the extension of the principle to cases where the magnetic field had a component normal to the surface, first with the addition of some unphysical~\cite{drevlak2005} or unnecessary sources~\cite{lazerson2012}, and then in the most general and mathematically-rigorous way~\cite{hanson2015}.

Envisaged applications of GSA include, above all, the processing of signals measured by experimental magnetic diagnostics to exclude contributions due to external sources, and the coupling with downstream equilibrium reconstructions that might therefore benefit from a higher degree of accuracy, a shortened runtime, or fewer degrees of freedom.

The present paper is organized as follows.
First, in Section~\ref{sec:math}, the VCP equation for GSA is introduced and its significance thoroughly discussed.
Right after that, Section~\ref{sec:context} further establishes the context of GSA and the classic VCP applications.
Then, in Section~\ref{sec:physics}, the application to tokamak experiments is presented, and then specifically applied to SPARC and its PRD simulation.
Furthermore, in Section~\ref{sec:code}, the novel computation code developed to tackle GSA is introduced and briefly summarized.
In Section~\ref{sec:results}, the main predictions of GSA are presented and the results are discussed.
Finally, in Section~\ref{sec:conclusions}, the most important features of the algorithm are highlighted once again, together with closing remarks.

\section{Mathematics}
\label{sec:math}

The most appropriate treatise of GSA comes from Hanson's 2015 work~\cite{hanson2015} introducing the most general expression of the Virtual-Casing Principle~(VCP).
In the following we discuss the \textit{magnetostatic} application of such generalized VCP, so that further terms deriving from non-solenoidal vector fields or non-stationary currents are discarded.
The interested reader should follow the complete derivation in the original work, as only the main result is presented here.

\subsection{Derivation}
\label{subsec:derivation}

In the short paper, Hanson arrives at a general two-dimensional surface integral which, under the correct geometric assumptions, equivalently represents a Biot-Savart-like three-dimensional volume integration for the magnetic contribution of sources internal to a bounded volume.
The pivotal step in the derivation of the integral is the choice of exploiting known vector identities to shift derivations from the magnetic curl over to the geometric factors that depend exclusively on the surface and the observation point.

Let us break the generality of the VCP and focus on the magnetostatic field $\mathbf{B}$, which is a divergence-free vector field whose curl is proportional to the current density $\mathbf{J}$ following Amp\`ere's Law:
\begin{align}
    \nabla \cdot \mathbf{B} &= 0 , \label{eq:divb}
    \\
    \nabla \times \mathbf{B} &= \mu_0 \ \mathbf{J} . \label{eq:curlb}
\end{align}
Apart from the geometric considerations introduced in the continuation of this work, the reduced system represented by Equations~\eqref{eq:divb} and~\eqref{eq:curlb} is the only physical assumption required by GSA.

Let us introduce a volume $V$ bounded by a surface $S=\partial V$, with outward normal unit vector $\mathbf{\hat{n}}$.
No assumptions are required for the configuration of currents flowing in space.
The \textit{observation} point $\mathbf{r}$ lies anywhere in space, whereas the \textit{integration} point $\mathbf{r'}$ lies in the volume $V$ or on its surface $S$.
Then, $\mathbf{a}$ represents the geometric distance factor due to the inverse-square law:
\begin{equation}
    \label{eq:a}
    \mathbf{a}(\mathbf{r},\mathbf{r'}) = \frac{\mathbf{r}-\mathbf{r'}}{|\mathbf{r}-\mathbf{r'}|^3} .
\end{equation}
Evidently $\mathbf{a}$ is singular where $\mathbf{r} = \mathbf{r'}$, and indeed the integration of the corresponding Dirac delta leads to a constant, $k$, which takes different values depending on whether $\mathbf{r}$ is internal, superficial, or external to the volume $V$:
\begin{equation}
    k =
    \begin{cases}
        1, & \mathbf{r} \in V
        \\
        1/2, & \mathbf{r} \in \partial V
        \\
        0. & \mathbf{r} \notin V, \partial V
    \end{cases}
\end{equation}
The main equation of the generalized VCP for the magnetostatic field, Equation~(19) in the original work, is finally given for $\mathbf{B}_V$, which represents the magnetic field due to sources internal to the volume $V$, as:
\begin{align}
   \mathbf{B}_V
   &=
   \frac{\mu_0}{4 \pi} \int_V
   \mathbf{J}(\mathbf{r'}) \times
   \frac{(\mathbf{r}-\mathbf{r'})}
   {|\mathbf{r}-\mathbf{r'}|^3}
   \mathrm{d}^3\mathbf{r'} \label{eq:biotsavart}
   \\[1em]
   &=
   \frac{1}{4 \pi} \oint_{\partial V} [
   \mathbf{a} \, (\mathbf{\hat{n}} \cdot \mathbf{B})
   - \mathbf{a} \times (\mathbf{\hat{n}} \times \mathbf{B})
   ] \, \mathrm{d}^2 \mathbf{r'}
   \, + \, k \, \mathbf{B} . \label{eq:vcp}
\end{align}

\subsection{Discussion}
\label{subsec:discussion}

As an important distinction, please note that the magnetic fields $\mathbf{B}_V$ and $k \, \mathbf{B}$, which appear \textit{outside} of the integral, depend on the \textit{observation} point $\mathbf{r}$, while the magnetic field in the scalar and vector product \textit{inside} the integral depends on the \textit{integration} point $\mathbf{r'}$, that is, the dummy integration variable.

Indeed, as long as the observation point $\mathbf{r}$ is external to the surface $S$, the Dirac delta factor $k$ drops out and the VCP is a straightforward equivalence of the Biot-Savart volume integral~\eqref{eq:biotsavart} to the surface integral in~\eqref{eq:vcp}.
Even more importantly, in such case the dependence of the magnetic field $\mathbf{B}$ from either the observation or the integration point is fully separated in the two sides of the equation, and therefore the knowledge of $\mathbf{B}(\mathbf{r'})$ on the surface $S$ is alone sufficient to fully yield $\mathbf{B}_V(\mathbf{r}) \equiv \mathbf{B}_\mathrm{in}$.

Vice versa, for observation points inside the volume $V$, the presence of the non-zero $k$ term `flips' the meaning of the surface integration, so that the addition of the total magnetic field $\mathbf{B}$ is needed to re-obtain the field due to internal sources.
In such case the knowledge of $\mathbf{B}(\mathbf{r'})$ on the surface $S$ actually yields $-\mathbf{B}_\mathrm{out}(\mathbf{r})$, instead, and is not sufficient to yield $\mathbf{B}_V(\mathbf{r})$.

Replacing the integrand of the surface integration with $\bigstar$, and separating the surface integral explicitly, the above considerations can be schematically expressed as:
\begin{equation}
    \label{eq:gsa}
    \oint_{\partial V} \bigstar \, \mathrm{d}^2 \mathbf{r'} =
    \begin{cases}
        - \mathbf{B}_\mathrm{out} , & \mathbf{r} \in V
        \\[1em]
        \displaystyle \frac{\mathbf{B}_\mathrm{in} - \mathbf{B}_\mathrm{out}}{2} , & \mathbf{r} \in \partial V
        \\[1em]
       \mathbf{B}_\mathrm{in} . & \mathbf{r} \notin V, \partial V
    \end{cases}
\end{equation}
Equation~\eqref{eq:gsa} represents the intuitive fact that, although the surface integral on $S$ can always be computed, its physical interpretation depends on the position of $\mathbf{r}$ relative to $S$.
As a consequence, in the following the observation point is always taken to be external to the surface $S$, so that the surface integration immediately yields the magnetic field at any \textit{external} point produced by \textit{internal} currents, without requiring knowledge of the external currents.

\section{Context}
\label{sec:context}

As a brief aside, let us dedicate this section to the similarities and differences of the GSA with the more established VCP.

In its most intuitive form, the VCP consists in the equivalence of a three-dimensional volume integral to a two-dimensional surface integral -- above all, such a promise of \textit{reduced dimensionality} is what appeals most to computational MHD applications.

Furthermore, our specific application enjoys the \textit{complete separation} of internal and external fields under the appropriate conditions.
The possibility of a full separation is intrinsic to the VCP and the underlying mathematics and vector-field physics, yet is rarely applied to the extent required by GSA, with usually limited information being extracted through boundary computations.

Finally, our approach does not depend on any specific model for any source anywhere, including the plasma, the external coils, or passive currents in the support structures.
We consider this absence of requirements on any current source or its modeling as the algorithm being \textit{model-agnostic}, to differentiate from other methods which do apply a certain set of assumptions on either the current sources or their models.
Fundamental differences can be found between the work detailed in this manuscript and the classic and well-established approaches which rely on the VCP.

\subsection{Coil design}

Coil design is a typical workflow which leverages the VCP to obtain a magnetic ``separation'', which might be named more appropriately a magnetic \textit{boundary-condition matching}.
The solution to the fixed-boundary problem -- which yields the total magnetic field in the internal region up until the boundary of the computation, that is, the outermost flux surface -- is usually obtained \textit{before} focusing on the design of the external magnetic field.
The issue of finding a set of external coils able to support the desired equilibrium is then framed as a problem of matching the magnetic boundary conditions at the domain boundary.
The assumption that the boundary is a flux surface requires that the \textit{normal} component of the external field is equal and opposite to that of the internal field.
The latter can be computed normally through a Biot-Savart volume integration of known quantities, but the VCP can be applied to yield a more efficient surface-integration formula.
This approach is model-dependent and provides a boundary condition for further models dealing with coil design and optimization in the external domain.

\subsection{Field extension}

Another application of the VCP involves ``vacuum-field extension'', namely the computation of a custom scalar potential satisfying Laplace's equation in order to extend the field in the external domain.
The irrotational field produced by such potential is typically needed to enforce the cancellation of the normal component of the total magnetic field on any chosen surface, which need not be a flux surface.
The solution of the resulting integral equation on the boundary allows then a similar computation to be carried out outside, yielding a solution to the exterior Neumann-boundary-condition problem.
Such solution to Laplace's equation is a convenient scalar potential that, through derivation, leads to an efficient computation of the magnetic field in the external region as due to internal sources.
This approach introduces, computes, and exploits an explicit scalar potential, while the GSA approach deals with total and internal fields only, instead.
Any comparison of the mathematical convenience or numerical efficiency of the GSA computation of a 2D surface integral with the solution to Laplace's equation -- either as a second-order Partial Differential Equation or as an integral equation -- is beyond the scope of this paper.

\subsection{Singularity integration}

As a final remark, let us acknowledge the considerable work carried out in order to efficiently solve the singularity function appearing in integral equations such as the VCP.
Specifically, recent work~\cite{malhotra2020} mentioned the difference between \textit{on-surface} and \textit{off-surface} evaluations, and then focused on the former by proposing a numerically-advanced method to tackle the singularity of the kernel.
The present work focuses on off-surface evaluations, instead, and thus does not rely on recent advanced numerical tools but implements a straightforward two-dimensional surface integration.

\subsection{Future work}

This simplified numerical approach is justified by two separate facts.

On one hand, the generalized VCP equation that underlies our work \textit{does not present} any singularity at all for off-surface evaluations, since the derivative of the singular term has been already integrated to yield the constant $k$, and since one can choose an appropriate, but finite, $\mathbf{r}-\mathbf{r'}$.
Even if future versions of the code were to tackle the on-surface evaluation and accurately and efficiently implemented a singularity-subtraction technique, the external magnetic field -- including passive currents on the vessel -- would then be required according to the \textit{caveat} of Section~\ref{subsec:discussion} and as such the purpose of the algorithm would be defied.

On the other hand, the primary goal of the continued development of this code is to directly ingest data coming from magnetic diagnostics, and as such will deal with real-world data with systematic and random errors, not with unlimited accuracy as it usually happens for mathematical identities and when dealing with advanced double-precision numerical methods.
Initial rough estimates for the standard error expected for SPARC's magnetic diagnostics are approximately 0.5\%, and as such outweigh any small-order inaccuracy in the numerical integration.

Indeed, ongoing and future work on GSA involves a paradigm shift in terms of input quantities, leaping from simulation data with perfect knowledge to measured data with partial knowledge and careful interpolation.
Nonetheless, this is outside the scope of the current exploratory study of GSA and will be the subject of a future publication, in the spirit of progressively building towards complexity.

\section{Physics}
\label{sec:physics}

First, let us discuss the similarities between the mathematics of GSA and the physics of tokamaks in general, and only then tackle the application to SPARC and its specifics.

\subsection{Tokamaks}

Magnetically-confined fusion experiments offer a natural application of Gauss' Separation Algorithm.
To begin, the vacuum vessel walls can be intuitively recognized as the surface $S$, which bounds the volume $V$, that is, the toroidal chamber.
In reality, the surface $S$ will correspond to the virtual surface defined by the `envelope' of the positions of the magnetic diagnostics that measure the total field and constitute the source of the magnetic information.
Indeed, the experimental designs of the machines invariably plan for a full set of magnetic diagnostics on the inner side of the vessel walls, facing the plasma, in order to inform feedback control algorithms, equilibrium reconstruction codes, and other tools.
Moreover, currents flowing on external or internal coils are pre-programmed, feedback-controlled, or otherwise known and measured easily.
Above all, the complex plasma physics and behavior is often entirely contained \textit{within} the vacuum vessel walls, so that the interesting internal plasma phenomenology can be naturally decoupled from the external physics through a GSA approach.

This simplified interpretation of the intertwined magnetic contributions to the total field is often complicated by the presence of eddy currents flowing in the passive structures of the machine, or halo currents flowing in the colder scrape-off layer of the plasma and the plasma-facing components.
Such passive or induced currents are in general inherently three-dimensional, most notably if flowing in the vessel that is riddled with ports and cuts.
Furthermore, these passive contributions might be negligible during parts of the discharge, yet quite significant during other phases.
It is known that the accuracy of equilibrium reconstructions is possibly hindered by this added layer of physical complexity, which needs to be properly taken into account~\cite{moret2015}.

Since GSA separates internal from external contributions with respect to the specified surface, that is, the inner wall of the vacuum vessel, it follows that GSA allows the separation of \textit{external} coils and vessel eddy currents from any \textit{internal} coil, passive structure, and plasma contribution.
Still, the GSA approach will \textit{not} be able to separate the contributions of internal passive structures or coils from those produced by the plasma -- for such a workflow further physics-informed models are needed.
In other words, external driven or induced currents are systematically separated from the rest, while internal driven or induced currents are `bundled' together so that the interpretation of the computed field due to \textit{internal} sources might require special care.

As a final consideration, let us stress that the GSA approach is fully three-dimensional and can leverage any source of magnetic information on the integration surface, including virtual or experimental diagnostics, or upstream simulation codes.

\subsection{SPARC}
\label{subsec:sparc}

SPARC, under construction by Commonwealth Fusion Systems in Devens, Massachusetts, USA, is a high-field tokamak predicted to achieve $Q=11$ using deuterium and tritium fuel~\cite{creely2020}.
The following work leverages the simulations of the SPARC Primary Reference Discharge~(PRD) performed at the end of 2021 using the Tokamak Simulation Code~(TSC)~\cite{jardin1986}, and thus applies to an outdated SPARC design.
As an example, the shape of the vacuum vessel received a minor update in 2022, and the vertical stability plates were dropped from the design.

\subsubsection{Structures}

\begin{figure}[b]
\centering
\fbox{\includegraphics[width=.85\columnwidth]{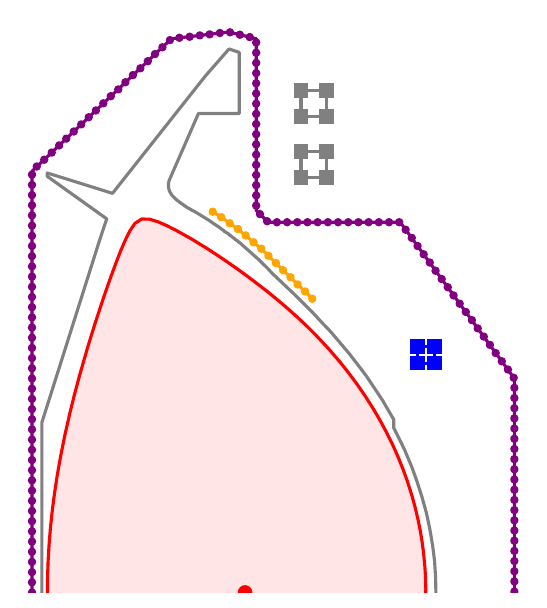}}
\caption{Simplified upper poloidal cross-section of SPARC, with the plasma region, boundary and magnetic axis~(red), limiter wall~(gray), upper VSP~(orange), upper VSC~(blue), vacuum vessel~(purple), and upper divertor coils~(gray). Within TSC, structures are either active coils~(squares) or passive conductors consisting of individual filaments~(circles).}
\label{fig:sparc}
\end{figure}

Figure~\ref{fig:sparc} shows a simplified cross-section of SPARC.
As already mentioned, any contribution from  \textit{external} conductors will be mathematically removed from the final result by the nature of the GSA approach, and can thus be safely disregarded when describing the machine.
Out of the external coils, only the upper divertor coils are shown.

On the other hand, the TSC simulation includes internal components which do contribute to the final result for the internal magnetic field, namely the Vertical Stability Coils~(VSCs) and the Vertical Stability Plates~(VSPs, no longer part of the design), both of which consist of an upper and a lower structure.
While the VSCs are actively fed via control algorithms for vertical stability purposes, passive unknown currents flow in the VSPs, which are modeled in TSC using a multi-filamentary approach.
Throughout the following, \textit{internal sources} will thus include the plasma, the VSCs, and the VSPs.

Finally, eddy currents are routinely induced on the vacuum vessel~(VES) itself, and have to be considered either external or superficial depending on how the integration surface for GSA is set up.
The TSC modeling of such passive currents once again consists in a multi-filamentary approach analogous to the one applied for the VSPs and the VSCs.

\subsubsection{Discharge}
\label{subsubsec:discharge}

\begin{figure}[b]
\centering
\fbox{\includegraphics[width=1\columnwidth]{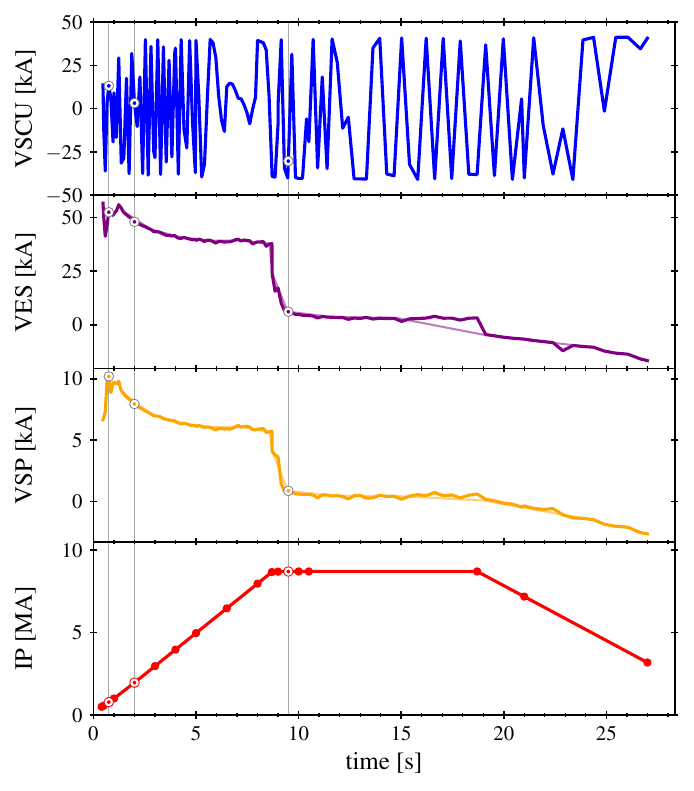}}
\caption{Time traces of the discharge currents, with gray vertical lines highlighting the (a)~early and (b)~late ramp-up, and the (c)~flat-top. The \textit{group} current is shown for the upper VSC~(blue), for the vessel currents~(VES, purple), for the VSP currents~(orange), and for the plasma current~(IP, red). For the passive currents~(VES, VSP), a lighter line is also showing the mostly-overlapping total \textit{individual} filament current.}
\label{fig:currents}
\end{figure}

The PRD discharge can be characterized by the time traces of currents in the plasma, and in the coils and passive structures.
Figure~\ref{fig:currents} shows the \textit{group} currents flowing in the various structures and in the plasma region.
 For the passive structures~(VES, VSPs), a lighter line is also showing the total \textit{individual} currents flowing in the multiple filaments.
The group and individual currents represent different information sources for the currents flowing in the passive structures, the former being \textit{uniformly} shared among the various filaments of the passive structure, and with the latter varying \textit{poloidally} between the filaments.
Although these two pieces of information are derived from the same internal model employed by TSC, they are output with different time bases and numerical precision, and so they will be treated as almost different models in the continuation of this work.
Indeed, the sparse time base of the individual currents does not cover the initial non-linear ramp-up phase of the passive currents and starts at a later time, $t$ = 1.43 s, before which only the group current information is available for the passive structures.

Out of the whole discharge, let us single out three representative instants of time, during the (a)~early and (b)~late ramp-up, and the (c)~flat-top, as summarized in Table~\ref{tab:stats}.
The choice of two ramp-up instants is a consequence of the absence of early information about the individual currents, together with the fact that a GSA approach is nonetheless applicable to such interesting early moments by using the group current, instead.

\begin{table}[!t]
\centering
\begin{tabular}{ccccc}
\hline \hline
& (a)~Early & (b)~Late & (c)~Flat & Unit \\
\hline
time &  \tramp &   \tlate &   \tflat & s \\
VSC & \vscramp & \vsclate & \vscflat & kA \\
VES & \vesramp & \veslate & \vesflat & kA \\
VSP & \vspramp & \vsplate & \vspflat & kA \\
IP &   \ipramp &  \iplate &  \ipflat & MA \\
\hline \hline
\end{tabular}
\caption{For the three representative moments of the discharge, (a)~early and (b)~late ramp-up, and (c)~flat-top, the total currents of vertical stability coils~(VSC), vessel eddy currents~(VES), vertical stability plates~(VSP), and plasma current~(IP) are shown.}
\label{tab:stats}
\end{table}

\subsubsection{Early ramp-up}

\begin{figure*}[!bt]
\centering
\fbox{\includegraphics[width=.33\textwidth]{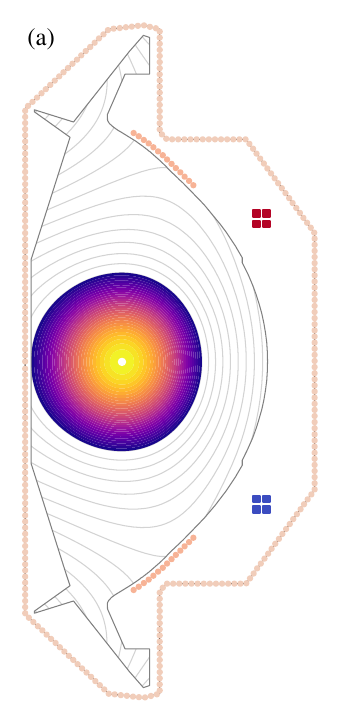}%
\includegraphics[width=.33\textwidth]{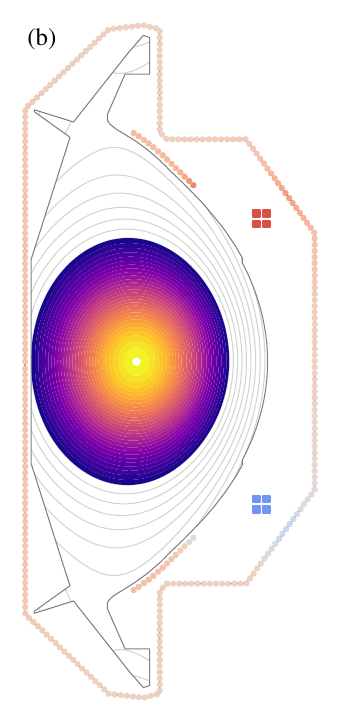}%
\includegraphics[width=.33\textwidth]{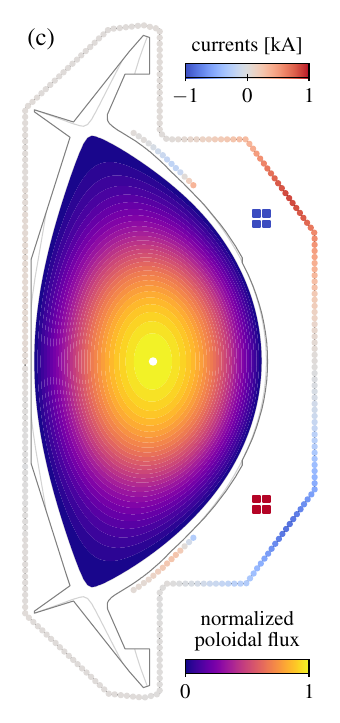}}%
\caption{Equilibria during the %
(a)~early ramp-up at $t$ = \tramp\ s, %
(b)~late ramp-up at $t$ = \tlate\ s, %
(c)~flat-top at $t$ = \tflat\ s. %
The poloidal flux is normalized from magnetic boundary to axis (bottom colorbar), while the internal currents are shown as individual filaments (top colorbar). Contours of the flux are also shown in the region external to the LCFS~(gray). In~(a) the group current on each structure is spread among all filaments, while in~(b-c) the individual-filament currents are used and shown.}
\label{fig:slices}
\end{figure*}

At $t$ = \tramp\ s, the plasma current is ramping up, and the plasma column is circular and inner-wall limited.
Passive currents on the vessel and VSPs, of the order of tens of kA, might complicate the convergence of equilibrium reconstruction codes.
Figure~\ref{fig:slices}a shows the magnetic flux surfaces for the equilibrium during the early ramp-up, and the group currents flowing \textit{uniformly} in the passive structures.

\subsubsection{Late ramp-up}

At $t$ = \tlate\ s, the plasma current is still ramping up, but the plasma column is now more elongated.
Passive currents on the vessel and VSPs remain roughly of the same order of magnitude as the early ramp-up.
Figure~\ref{fig:slices}b shows the magnetic flux surfaces for the equilibrium during the late ramp-up, and the \textit{poloidally-localized} filament currents flowing in the passive structures.
At this time, the poloidal localization of the eddy currents on the vessel is weak, as most filaments carry a low but positive current.

\subsubsection{Flat-top}

At $t$ = \tflat\ s, the plasma reached its flat-top phase with a double-null up-down-symmetric shape, and the induced currents are now much smaller, of the order of a few kA.
The VSCs are still being fed with tens of kA for vertical stability control, and indeed at this time the maximum current is flowing through the upper and lower VSC, with opposite polarity.
Figure~\ref{fig:slices}c shows the magnetic flux surfaces for the equilibrium during the flat-top, and the \textit{poloidally-localized} filament currents flowing in the passive structures.
With respect to the late ramp-up, at this time the poloidal localization of the eddy currents on the vessel is more apparent, as most filaments carry negligible currents while only a few poloidal sections next to the VSCs exhibit kA currents.
Even then, large currents on some poloidal sections might be unrealistic as they would flow in regions where ports exist on the physical vacuum vessel.
Still, the inclusion of eddy currents is needed for a correct characterization of the total magnetic field, and any validation of the wall model used by TSC is outside the scope of this work.

\subsection{ITER}

As a side note let us mention that ITER, in addition to the usual in-vessel magnetic diagnostics, will also provide a set of ex-vessel diagnostics~\cite{vayakis2012}.
In such a case a `dual' GSA approach might be used to extract three different regions from the magnetic measurements~\cite{sweeney2019}: a fully internal region, a fully external region, but also an intermediate region mostly containing vessel currents.

\section{Code}
\label{sec:code}

A novel computation code -- named \texttt{GaSATO} for \underline{\textbf{Ga}}uss' \underline{\textbf S}eparation \underline{\textbf A}lgorithm for \underline{\textbf T}okamak \underline{\textbf O}perations -- was developed in order to apply the GSA approach to the SPARC~PRD simulation.
The code, written in Python and leveraging established numerical libraries~\cite{harris2020, ortner2020, hoyer2017}, runs on a consumer laptop and is capable of executing in a few tens of seconds in its current unoptimized form.
Future plans include the full parallelization of the workflow using cutting-edge frameworks.

\subsection{Inputs}
\label{subsec:inputs}

In the most general terms, the code reads magnetic data as input, and provides magnetic data as output.
Although the final production-ready implementation of the code will certainly differ, the present code accepts TSC data as input and solves the GSA problem for a given set of parameters on a given number of points.

The upstream equilibrium and simulation code, TSC~\cite{jardin1986}, adopts an assumption of \textit{axisymmetry}.
As anticipated, the main envisaged application of GSA consists in preprocessing experimental magnetic data to enhance downstream equilibrium reconstructions.
Any improvement of such axisymmetric workflows will have an immediate and lasting impact on the operation of SPARC, either as a control-room tool or as a real-time-control algorithm, particularly when significant passive currents might occurr like during ramp-up.
Fully 3D simulations will be investigated in the future.

For the purposes of applying GSA, the TSC simulation consists of a time-based series of equilibrium files together with a set of matrices describing the position and currents of the internal coils and passive structures, which are described as individual filaments.
The time-dependent series of data represents all moments of the discharge, including ramp-up, flat-top, and ramp-down.
In general, the time base of the series of equilibria differs from that of the current matrices.
Precedence was accorded to the time base of the equilibria, since time interpolation of one-dimensional signals like currents is preferable to time interpolation of two-dimensional quantities like the plasma poloidal flux, which already require an interpolation in space between the $(R,Z)$ grid points.

As already mentioned in Section~\ref{subsubsec:discharge}, both a \textit{group} current and a set of \textit{individual} currents are provided for the passive structures, with different time bases.
During the early ramp-up phase the information about the individual currents is not yet available, and thus only the group current can be applied.

In the GSA approach only the \textit{total} magnetic field $\mathbf{B}$ on the integration surface $S$ is needed to compute the surface integral from Equation~\eqref{eq:vcp}, and thus find the magnetic contribution due to internal sources.
The knowledge of the distribution of internal currents is only needed in order to allow a quantification of the agreement of the `computed' internal field, that is, the result of GSA, with the `modeled' internal field, that is, the magnetic field arising from internal sources according to TSC.

\subsection{Surfaces}

A GSA simulation is executed for each of the provided equilibria during the discharge.
As designed, the integration surface $S$ is a slightly `imploded' version of the vacuum vessel, and the total field $\mathbf{B}$ is computed on $S$ through the equilibrium files.
At this stage the integration surface can be chosen at will, since we are dealing with simulation data through equilibrium files that provide the magnetic field everywhere in the vacuum chamber, but in reality in future experimental applications the integration surface will be decided by the `envelope' of the position of the magnetic diagnostics providing the actual magnetic field information.
Although the equilibria under examination are axisymmetric in nature, the surface integration still spans the whole surface of a three-dimensional torus-like structure, obtained by rotating $S$ around the $z$ axis.

\begin{figure}[b]
\centering
\fbox{\includegraphics[width=1\columnwidth]{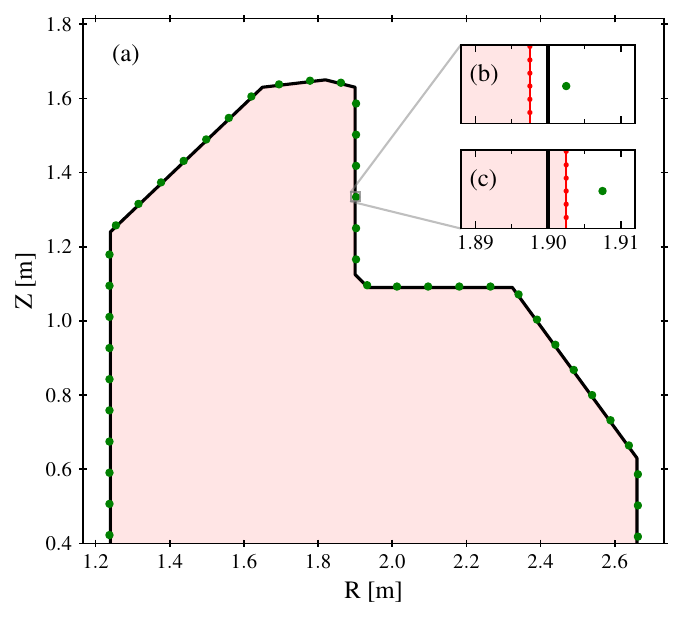}}
\caption{Simplified upper poloidal cross section of the surfaces of importance in the GSA approach. (a) A zoomed-out view of the upper poloidal cross section, with the observation points~(green), the vacuum vessel~(black), and the integration surface and internal region~(red). Two zoomed-in boxes show the millimetric distances between the integration surface~(red), the vacuum vessel~(black), and one of the slightly external observation points~(green) in both the (b)~design in-vessel case, and the (c)~ex-vessel case.}
\label{fig:surfaces}
\end{figure}

At the same time, the observation path $T$, which is simply a collection of $N_T$ points on which GSA is applied, is a slightly `exploded' version of the vacuum vessel outline, so that the distance between $S$ and $T$ is 5 mm and any divergence of the singular $\mathbf{a}$ term from Equation~\eqref{eq:a} is avoided.
At present, again due to the axisymmetric nature of the equilibrium fields, $T$ lies on the $\phi = 0$ plane.
Since the observation path is external to the integration surface according to the \textit{caveat} of Section~\ref{subsec:discussion}, the contribution of internal sources to the magnetic field on $T$ is completely defined by the knowledge of $S$, and $\mathbf{B}$ on $S$.
Figures~\ref{fig:surfaces}a-b show a general and magnified view of these surfaces.

The torus-like 3D structure is further chunked into several portions with varying degrees of fidelity in order to speed up computations.
Specifically, the portion of surface closer to the observation points, with $|\mathbf{r}-\mathbf{r'}|$ no less than 5 mm, is subdivided with a fine interpolation grid, the portion of surface on the opposite side of the torus, with roughly $|\mathbf{r}-\mathbf{r'}| \approx 2 R_0 \approx 4 \ \mathrm{m}$, is subdivided with a coarse interpolation grid, and the intermediate portions are subdividided with medium grids.
This chunked approach is made possible by having $T$ lying on the $\phi = 0$ plane, not by having axisymmetric equilibrium data as input.

Even though the design case for SPARC is that of \textit{in-vessel} diagnostics, another set of simulations was executed with an \textit{ex-vessel} design, where diagnostics mounted on the exterior vacuum vessel wall ideally provide information on a slightly external surface so that the vessel is included in the GSA region, as shown in Figure~\ref{fig:surfaces}c.
Albeit non-physical for SPARC, such a \textit{Gedankenexperiment} helps to establish the robustness of the GSA algorithm in case the intense eddy currents flowing in the vessel were internal to the integration surface, not external, and `curled' in such a way to matter towards the computed internal field and possibly lead to destabilizing numerical effects.

\subsection{Metrics}

The GSA algorithm allows the computation of the magnetic field due to sources \textit{internal} to the surface $S$, on any chosen external point, only by knowing $\mathbf{B}$ on $S$ and by solving the numerical integral of Equation~\eqref{eq:vcp}.

An actual model for the plasma is exploited in order to provide a valuable comparison metric to the prediction of GSA.
Specifically, the plasma contribution is computed through a distributed-current model according to the source profiles from the TSC simulation, whereas the VSCs and VSPs are modeled with a multi-filamentary approach from their known and reconstructed currents.
The application of such established models for the internal current sources leads to a GSA-independent determination of the internal magnetic field on the observation path, $\mathbf{B}_\mathrm{int}$, and allows a quantification of the agreement of such quantity with the actual result of the GSA prediction, $\mathbf{B}_\mathrm{gsa}$.

A classic mean square error~(MSE) can then be introduced as:
\begin{equation}
    \mathrm{MSE} = \frac{1}{N_T} \sum_{N_T} \left| \mathbf{B}_\mathrm{gsa} - \mathbf{B}_\mathrm{int} \right| ^2 .
\end{equation}
It is found that the MSE varies significantly during the discharge, as does the relative magnitude of the reconstructed magnetic field.
A much better quantity is then a \textit{normalized} root mean square error~(NRMSE), which is normalized with respect to the maximum value of the norm of the internal magnetic field:
\begin{equation}
    \mathrm{NRMSE} = \frac{ \sqrt{\mathrm{MSE}} }{ \mathrm{max}\ |\mathbf{B}_\mathrm{int}| }.
\end{equation}
The adimensional NRMSE parameter is expressed as a percentage error.

As a further assessment of the robustness of the solution, it is appropriate to introduce a local, rather than averaged, metric like the maximum absolute error~(MaAE):
\begin{equation}
    \mathrm{MaAE} = \mathrm{max} \ | \mathbf{B}_\mathrm{gsa} - \mathbf{B}_\mathrm{int} | ,
\end{equation}
together with its normalized and adimensional equivalent:
\begin{equation}
    \mathrm{N MaAE} = \frac{\mathrm{max} \ | \mathbf{B}_\mathrm{gsa} - \mathbf{B}_\mathrm{int} |}{\mathrm{max} \ | \mathbf{B}_\mathrm{int} |} .
\end{equation}
The NMaAE allows to detect any localized and consistent error that might otherwise be smoothed out by the averaging operation, and thus complements the averaged quantification provided by the NMRSE.

In the following, the magnetic field due to internal sources at a given observation point is called `internal', and it is furthermore called `modeled' when arising from a classic multi-filament model for the plasma and the internal structures, or `computed' when obtained through the GSA approach.

\subsection{Parameters}

It is now appropriate to quickly summarize the various simulation parameters which set the different runs apart, as reported in Table~\ref{tab:params}.

The first parameter is the time during the discharge.
In the specific TSC upstream simulation, equilibrium files are provided for 17 time slices, including the three moments (a), (b), and (c) singled out in Section~\ref{subsec:sparc}.
The application of GSA through time allows the algorithm to be benchmarked during all moments of the discharge, including when internal coils or passive structures exhibit multi-kA currents.

The second parameter is the choice for the integration surface that can be either in the (d)~design case, with in-vessel diagnostics, or the (e)~ex-vessel case.
In the two design cases the eddy currents on the vessel are external, or internal, to the integration surface, respectively.
The application of GSA across the design cases tests the robustness of the algorithm with respect to currents which are quite close to the integration surface, and whose magnetic curl either cancels out in the mathematical integration, or yields significant contributions.

The third parameter is whether the passive structures are characterized using their group current, which is equally shared among each filament, or the individual currents, which are poloidally localized.
The application of GSA for both cases further tests the robustness of the algorithm with respect to significant internal currents, as the poloidal localization of the filament currents translates to much higher spikes of current quite close to the integration surface.
As previously mentioned, the results of the next section do not presume to validate the internal TSC model, but merely offer a comparison with the GSA solution.

\begin{table}[!bt]
\centering
\begin{tabular}{lrrr}
\hline
Parameter & & Cases \\
\hline
Time           & (a)~early     & (b)~late           & (c)~flat  \\
  \hfill $t$ = &     \tramp\ s &          \tlate\ s & \tflat\ s \\
Design         & (d)~in-vessel & (e)~ex-vessel      \\
               &       -2.5 mm &            +2.5 mm \\
Currents       & group         & individual         \\
               &         $I_g$ & $I_1, \cdots, I_M$ \\
\hline
\end{tabular}
\caption{Summary of the simulation parameters and their variation. The design parameter symbolizes the `implosion' or `explosion' of the vessel structure defining the integration surface. The currents for the passive structures can either be a group quantity, $I_g$, or an array of different values, $I_m$, for each filament $m=1, \cdots, M$.}
\label{tab:params}
\end{table}

\section{Results}
\label{sec:results}

\subsection{Discharge}

Let us first analyze the three singled-out moments of (a)~early and (b)~late ramp-up, and (c)~flat-top, before showing the results for the full discharge.

\begin{figure}[b]
\centering
\fbox{\includegraphics[width=1\columnwidth]{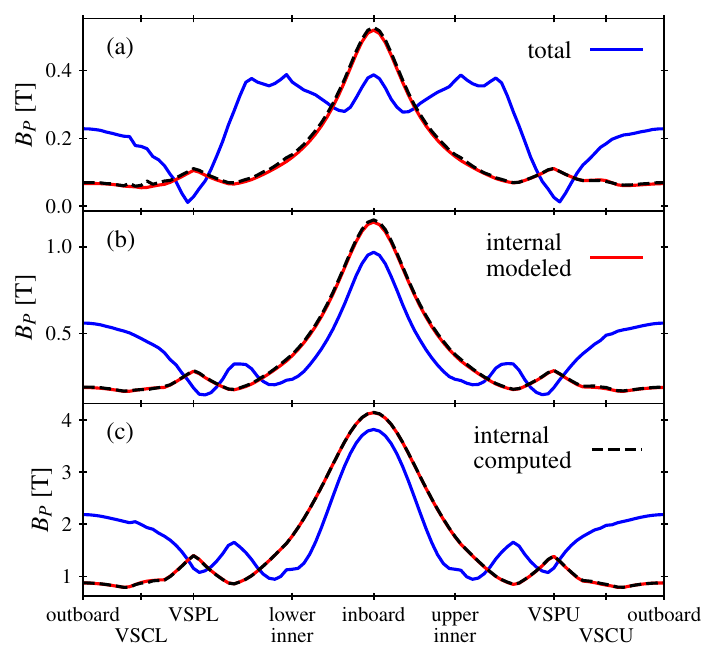}}
\caption{Profiles of the magnetic field along the observation path, annotated in terms of intuitive but irrelevant topological points along the vessel.
The panels correspond to the (a)~early and (b)~late ramp-up, and the (c)~flat-top case, with the total magnetic field~(blue), the modeled internal field~(red), and the computed internal field~(black, dashed).}
\label{fig:bpol}
\end{figure}

Figure~\ref{fig:bpol}a shows the total poloidal field, the modeled internal field, and the computed internal field for the (a)~early ramp-up case.
Although only a group current is available for the passive structures at this point, the NRMSE of approximately \rmseezpzramp \% suggests a good agreement, with the maximum error estimate of NMaAE roughly at \maaeezpzramp \%.
Similarly, Figure~\ref{fig:bpol}b shows the profiles of the fields for the (b)~late ramp-up case.
This is the first moment at which the individual current is applied to the filaments describing the passive structures.
The NRMSE of approximately \rmseezpulate \%, with NMaAE of \maaeezpulate \%, denotes an improved agreement.
Finally, Figure~\ref{fig:bpol}c shows the profiles in the (c)~flat-top case.
The NRMSE is now approximately \rmseezpuflat \%, with NMaAE of \maaeezpuflat \%, indicating an excellent agreement.

\begin{figure}[b]
\centering
\fbox{\includegraphics[width=1\columnwidth]{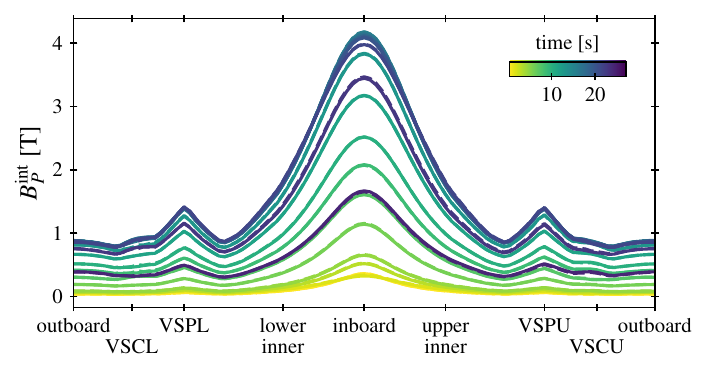}}
\caption{Profiles of the internal magnetic field along the observation path, annotated in terms of intuitive but irrelevant topological points along the vessel.
The plot shows the various time slices during the discharge~(top colorbar) in the (d)~design case, with the modeled internal field~(solid), and the mostly-overlapping computed internal field~(dashed).}
\label{fig:bint}
\end{figure}

Let us now focus on the simulations for the whole discharge.
Figure~\ref{fig:bint} shows the ensemble plot for the internal poloidal magnetic field for the (d)~design in-vessel case, where all reconstructed profiles exhibit a consistent agreement.
The (e)~ex-vessel case performs analogously so that its ensemble plot would not offer further insight into the numerics of GSA and is thus omitted.
Table~\ref{tab:nrmse} reports the resulting NRMSE values for the three singled-out moments in both design cases and for both current data sources.

\begin{table}[!bt]
\centering
\begin{tabular}{ccccc}
\hline
Design & Currents & (a)~Early & (b)~Late & (c)~Flat \\
\hline
(d)~in-ves & group  & \rmseezpzramp & \rmseezpzlate & \rmseezpzflat \\
           & single &         -     & \rmseezpulate & \rmseezpuflat \\
(e)~ex-ves & group  & \rmseeupzramp & \rmseeupzlate & \rmseeupzflat \\
           & single &         -     & \rmseeupulate & \rmseeupuflat \\
\hline
\hline
\end{tabular}
\caption{For the three representative moments of the discharge, (a)~early and (b)~late ramp-up, and (c)~flat-top, the resulting percentage NRMSE values are reported for the (d)~design in-vessel case and the (e)~ex-vessel case, and for the single~(individual) and group current data sources.}
\label{tab:nrmse}
\end{table}

\subsection{Ensembles}

\begin{figure*}[!bth]
\centering
\fbox{\includegraphics[width=1\textwidth]{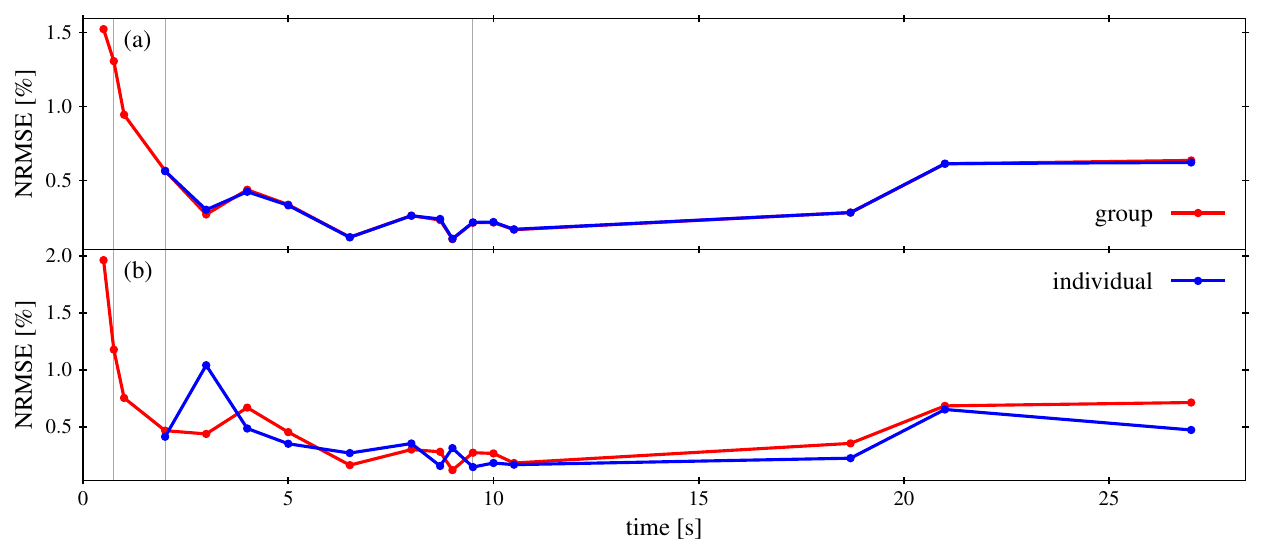}}
\caption{Time traces of the NRMSE parameter for the (a)~design in-vessel cases, and the (b)~ex-vessel cases, with gray vertical lines at the usual singled-out moments. Simulations with both individual-filament currents~(blue) and group currents~(red) are shown. The GSA prediction does not vary between individual-filament and group-current runs, but the modeled internal field does, and so does the resulting NRMSE parameter.}
\label{fig:nrmse}
\end{figure*}

Figure~\ref{fig:nrmse} shows the NRMSE traces for the four simulation ensembles, using both design cases (in-, and ex-vessel) and both current data source for the filaments on the passive structures (group, and individual).
Throughout all the simulations, the GSA-computed internal magnetic field agrees to a high degree of accuracy with its reference value modeled using a filamentary approach, having NRMSEs lower than a percentage point.
The specific value of the NRMSE parameter depends on several factors, the predominant being the errors due to the interpolation of the results of the TSC simulation, and the smallest being the number of observation points, or the specific grid size chosen for the integration.

Figure~\ref{fig:nrmse}a shows the NRMSE for the (d)~design in-vessel case, for both the individual and group currents.
The substantial overlap of the NRMSE curves for the group and individual currents indicates either that these two currents are in substantial agreement with each other, or that the contribution due to the passive structures~(VSPs) is relatively negligible.

Figure~\ref{fig:nrmse}b shows the NRMSE for the (e)~ex-vessel case, for both the individual and group currents.
Somewhat surprisingly, the group- and individual-current simulations appear to be quantitatively similar but qualitatively different, which might be explained by the absence of a common time base and the need for interpolation at each time step.
As already noted, the comparison of both TSC sources of information in the GSA simulations allows further insight into the numerics of the algorithm without intending to validate either one as the most appropriate current for the application of GSA.

The NMaAE for the ensemble values from the late ramp-up onwards, $t \geq \tlate\ s$, range from a minimum value of \maaelatemin \% to a maximum value of \maaelatemax \%, which is indicative of an excellent agreement as the biggest difference among the observation points is still only marginally worse than the average error provided by the NRMSE.
When considering the initial moments, as well, the NMaAE reaches a maximum of \maaefullmax \%, which is still reasonable if compared to the corresponding NRMSE of \rmsefullmax \%.

In any case, the GSA approach does confirm its worth as a computational method for magnetic separation based on information about the total magnetic field, and without further assumptions.

\section{Conclusions}
\label{sec:conclusions}

The present work introduced the Gauss' Separation Algorithm~(GSA) approach as a particular workflow built on top of a generalized Virtual-Casing principle.
Such solution is completely independent on any specific current-source modeling, with the exclusion of the reduced set of Maxwell's equations for the divergence and curl of $\mathbf{B}$, and is fully three-dimensional.
A novel MHD code named \texttt{GaSATO} was developed to tackle such computation, and its main results were presented.
The agreement of the magnetic field due to internal sources between the GSA computation and the corresponding multi-filamentary model is excellent in almost all phases of the discharge, with normalized root mean square errors of around 1\% in the first couple of instants, and mostly lower than 0.5\% afterwards, and with a normalized maximum absolute error slightly higher than that.
The biggest source of disagreement during the earliest times is certainly due to the missing current information and the relatively sparse timebase of the individual-current reference.

Although the VCP has been studied extensively, the GSA approach adds another application which promises a \textit{model-agnostic} and efficient way to reach the \textit{complete separation} of internal and external fields, and as such stands out and complements previous work.
Ongoing and future developments will introduce a magnetic projection onto a set of experimental magnetic diagnostics, an interpolation of the fields onto a fine grid, and the investigation of the robustness of the experimental setup for GSA separation in SPARC and other machines.

It is worth noting that the model-agnostic nature of the GSA approach does not preclude a combined workflow in which GSA is coupled with a specific model to test its validity, assess its robustness, extend its applicability, or simply improve its accuracy.
Indeed, envisaged applications of GSA include workflows to further improve equilibrium reconstruction during ramp-up or disruptions such as, but not limited to, vertical displacements events.
By exploiting GSA as a pre-processing step to filter out external currents from a set of magnetic measurements, subsequent downstream codes might benefit from fewer degrees of freedom for faster or more accurate equilibrium reconstruction, or even improved convergence.
Another workflow might include coupling GSA with verified models for the internal contributions, in order to study the unknown eddy currents on the passive structures.
Vice versa, a further example might include coupling GSA with a specific wall model in order to test the prediction of eddy currents under various assumptions.
Finally, on experimental devices capable of both in- and ex-vessel measurements, a dual GSA approach might provide separate predictions on internal and wall contributions.

As part of the optimization process of the code, GSA might be framed as a matrix multiplication problem.
By freezing the specific geometry, including the integration surface and the observation points, the problem might be tackled separately with a time-intensive compute-and-store approach, followed by a repeated but fast lookup-and-multiply problem.
Pushing the boundaries even further, a compromise on the numerical completeness of the integration in exchange for much faster solutions might allow algorithmic benefits compatible with real-time execution, and thus feedback-control applications.

\section*{Acknowledgments}
The first author wishes to thank FJ~Artola for suggesting the appropriate reference, and JD~Hanson for writing such an elegant and enlightening paper.

Work funded by Commonwealth Fusion Systems.

\section*{References}

\bibliography{../library}
\bibliographystyle{apalike}

\end{document}